\documentclass[aps,prl,reprint,nofootinbib,superscriptaddress]{revtex4-1}

\usepackage{graphicx}

\usepackage{color}\usepackage{enumerate}
\usepackage{amsmath}\usepackage{amssymb}\usepackage{stmaryrd}
\usepackage{multirow}
\usepackage{tcolorbox, accents}

\usepackage{longtable,booktabs}
\usepackage{hyperref}\usepackage{url}
\usepackage{color}\usepackage{enumerate} 
\usepackage{amsmath}\usepackage{amssymb}\usepackage{stmaryrd}
\usepackage{braket}
\usepackage{multirow}
\usepackage{float}
\usepackage[toc,page]{appendix}
\usepackage{longtable,booktabs}
 \usepackage{epsfig}
\usepackage{graphicx} 
\usepackage{feynmf}
\usepackage{psfrag} 
\usepackage{slashed}
\usepackage{amsmath,amssymb} 
\usepackage{colordvi} 
\usepackage{hyperref}
\usepackage{setspace}
\usepackage[all]{hypcap}
\usepackage{natbib}
\usepackage[normalem]{ulem}
\usepackage{subfigure}
\usepackage{dsfont}
\usepackage{amsfonts}
\usepackage{bbm}

\hypersetup{
    bookmarks=true,         
    unicode=false,          
    pdftoolbar=true,        
    pdfmenubar=true,        
    pdffitwindow=false,     
    pdfstartview={FitH},    
    pdftitle={My title},    
    pdfauthor={Author},     
    pdfsubject={Subject},   
    pdfcreator={Creator},   
    pdfproducer={Producer}, 
    pdfkeywords={keyword1} {key2} {key3}, 
    pdfnewwindow=true,      
    colorlinks=true,       
    linkcolor=cyan,          
    citecolor=purple,        
    filecolor=magenta,      
    urlcolor=black           
}
\usepackage{natbib} 
\def\be{\begin{equation}}
\def\bea{\begin{eqnarray}}
\def\eea{\end{eqnarray}}
\def\ee{\end{equation}}

\begin{document}
\def\AT#1{{\color{cyan}#1}}
\def\ATC#1{{\color{cyan}\xout{#1}}}
\def\FH#1{{\color{red}#1}}
\def\FHC#1{{\color{red}\xout{#1}}}

\title{An exactly soluble model for a fractionalized Weyl semimetal}

\author{Fabian Hotz}
\affiliation{Institute for Theoretical Physics, ETH Zurich, 8093 Z\"urich, Switzerland}

\author{Apoorv Tiwari}
\affiliation{Department of Physics, University of Zurich, 8057 Zurich, Switzerland}

\author{Oguz Turker}
\affiliation{Institut f\"ur Theoretische Physik and W\"urzburg-Dresden Cluster of Excellence ct.qmat, Technische Universit\"at Dresden, 01062 Dresden, Germany
}

\author{Tobias Meng}
\affiliation{Institut f\"ur Theoretische Physik and W\"urzburg-Dresden Cluster of Excellence ct.qmat, Technische Universit\"at Dresden, 01062 Dresden, Germany
}

\author{Ady Stern}
\affiliation{Department of Condensed Matter Physics, The Weizmann Institute of Science, Rehovot 76100, Israel}

\author{Maciej Koch-Janusz}
\affiliation{Institute for Theoretical Physics, ETH Zurich, 8093 Z\"urich, Switzerland}

\author{Titus Neupert}
\affiliation{Department of Physics, University of Zurich, 8057 Zurich, Switzerland}

\begin{abstract}
We construct an exactly solvable lattice model of a fractional Weyl semimetal (FWS). The low energy theory of this strongly interacting state is that of a Weyl semimetal built out of fractionally charged fermions. We show the existence of a universally quantized and fractional circular photogalvanic effect (CPGE) and a violation of the Wiedemann-Franz law in the system. Together with a spectral gap in the single-particle electronic Green's function they provide strong experimental signatures for this exotic gapless state of matter.
\end{abstract}

\maketitle


Electronic systems jointly described as topological fall in two conceptually distinct classes: short-range entangled, which can be transformed into a product state by means of local unitaries, and those characterized by long-range entanglement, which cannot. The former include symmetry protected topological states (SPTs) including topological insulators (TIs) or the Haldane phase \cite{PhysRevLett.50.1153}, and Weyl and Dirac semimetals. The latter class is composed of \emph{topologically ordered}~\cite{doi:10.1142/S0217979290000139} states characterized by topological ground state degeneracies and fractionalization of quantum numbers -- canonical examples thereof are fractional quantum Hall states (FQH). While many SPTs originate in the properties of the non-interacting band structure, and topological order crucially requires interactions, these subsets are not mutually exclusive: symmetry-enriched topological phases such as a fractional topological insulator (FTI) \cite{PhysRevLett.103.196803,PhysRevLett.105.246809,PhysRevB.84.235145} can arise due to the interplay of the two ingredients.



Recently, a lot of attention, both in theory and in experiment, has been devoted to Weyl semimetals (see references in \cite{2013arXiv1301.0330T,RevModPhys.90.015001}). Their deceptive simplicity -- a Weyl semimetal is essentially an accidental degeneracy in the band structure of a three dimensional solid \cite{PhysRev.52.365, Berry1985} -- hides rich phenomenology: Fermi arcs \cite{PhysRevB.83.205101}, negative magnetoresistance \cite{PhysRevB.88.104412,Burkov_2015} and the quantized circular photogalvanic effect (CPGE) \cite{de2017quantized}, among others. 
All of the above can be derived within the framework of noninteracting physics, but has been shown to be stable to interactions, which mostly move around and renormalize the Weyl cones~\cite{PhysRevB.85.045124,PhysRevLett.113.136402,PhysRevB.98.241102,PhysRevB.95.201102}. 
On the other hand, interactions may generate topological order which cannot be described within the framework of simple band theory, possibly creating non-trivial fractionalized counterparts to gapped noninteracting topological states, as demonstrated by the FTI. Indeed, fractionalized phases that possess Weyl nodes have been postulated at the boundary of a four-dimensional quantum Hall insulator~\cite{PhysRevB.94.155136}. Reference~\cite{PhysRevX.9.011039} constructs a three-dimensional (3D) topologically ordered Weyl phase, however, fractionalization is not present in its low-energy theory~\cite{abanin-misc}. 

In this work, we provide a realization of a fractionalized Weyl semimetal in 3D by constructing an exactly solvable lattice Hamiltonian of a topologically ordered Weyl phase, whose low-energy theory features fractionally charged excitations. We show that the CPGE, a second order response, is \emph{universally} quantized to fractional values, in clear contrast to standard Weyl semimetal. We also show a violation of the Wiedemann-Franz law and a gap in the single-particle electronic spectral function, which provide experimental signatures.

\begin{figure}[t]
 	\centering
 	\includegraphics[scale=0.31]{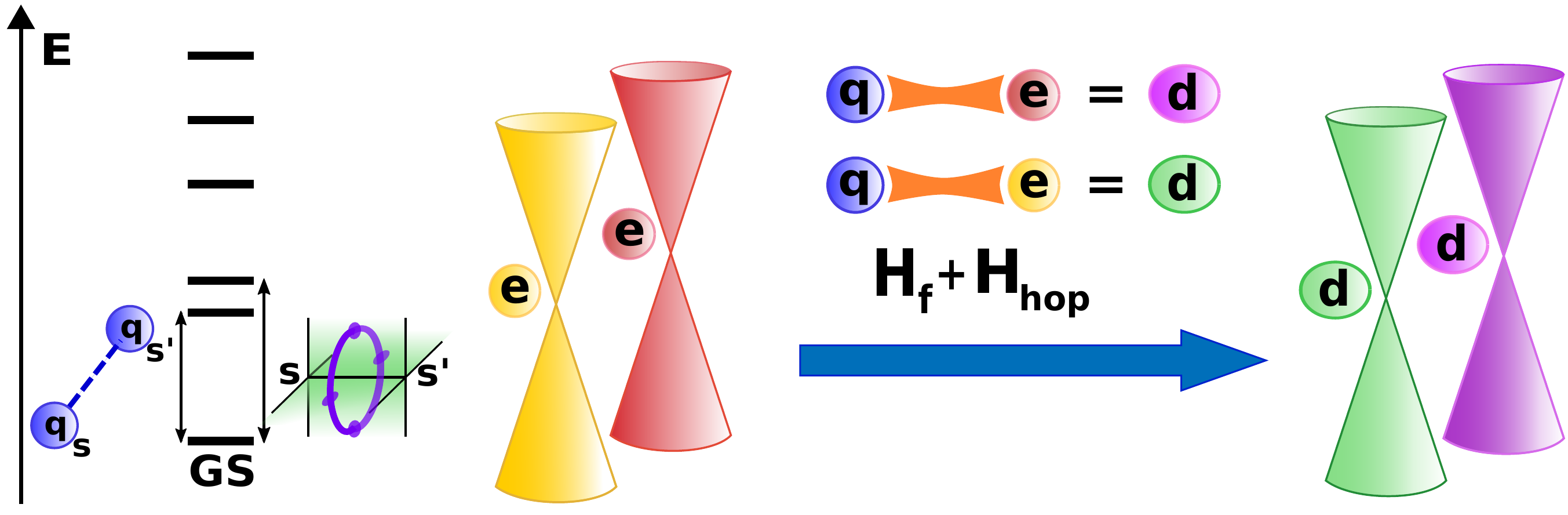}%
 	\caption{Our system is made up of a topologically ordered bosonic model coupled to electrons. Prior to coupling, the spectrum of the bosonic model is discrete, and the low energy theory is that of Weyl electrons $e$. The charging term in $\hat{H}_{\mathrm{f}}$ and the hopping $\hat{H}_{\mathrm{hop}}$ bind electrons to fractionally charged $q_s$ bosons. Their composite $d$ inherits the Weyl dispersion, while the original electrons become gapped [see Fig.(\ref{fig:spectral})]
 	}
 	\label{fig:cones}
 \end{figure}

\paragraph{The solvable model --} Our construction is based on a flexible blueprint for an exactly solvable lattice model of a $\mathbb{Z}_m$ lattice gauge theory with conserved U(1) charge, which some of us have previously introduced to study symmetry-enriched topological phases (in particular FTIs) and their properties in 2D and 3D \cite{PhysRevB.84.235145,koch2013exactly,PhysRevB.95.205110}. 
It proceeds in two stages: first a bosonic model on a lattice is introduced, realizing a discrete lattice gauge theory, similar to a generalized toric code \cite{kitaev2003fault}, albeit with the crucial distinction of having conserved boson number. The bosons carry a charge of $2e$ and can be thought of as paired electrons. This model features bosonic quasiparticles with fractional charge. In the second step, free electrons are added to the lattice and coupled to the bosonic quasiparticles. The band structure of the electrons encoded in the hopping amplitudes is chosen to be that of a Weyl semimetal; throughout this procedure the model remains exactly solvable. We show that the low energy physics of the model can be mapped to a Weyl semimetal built out of fermions of \emph{fractional} electric charge (see Fig.~\ref{fig:cones}), and analyze the experimental consequences. Below we describe the model in more detail; further information about this class of constructions can be found in Supplementary Materials (SM) and Refs.~\cite{PhysRevB.84.235145,koch2013exactly,PhysRevB.95.205110}. 
 
\begin{figure}[t]
	\centering
	\includegraphics[scale=0.43]{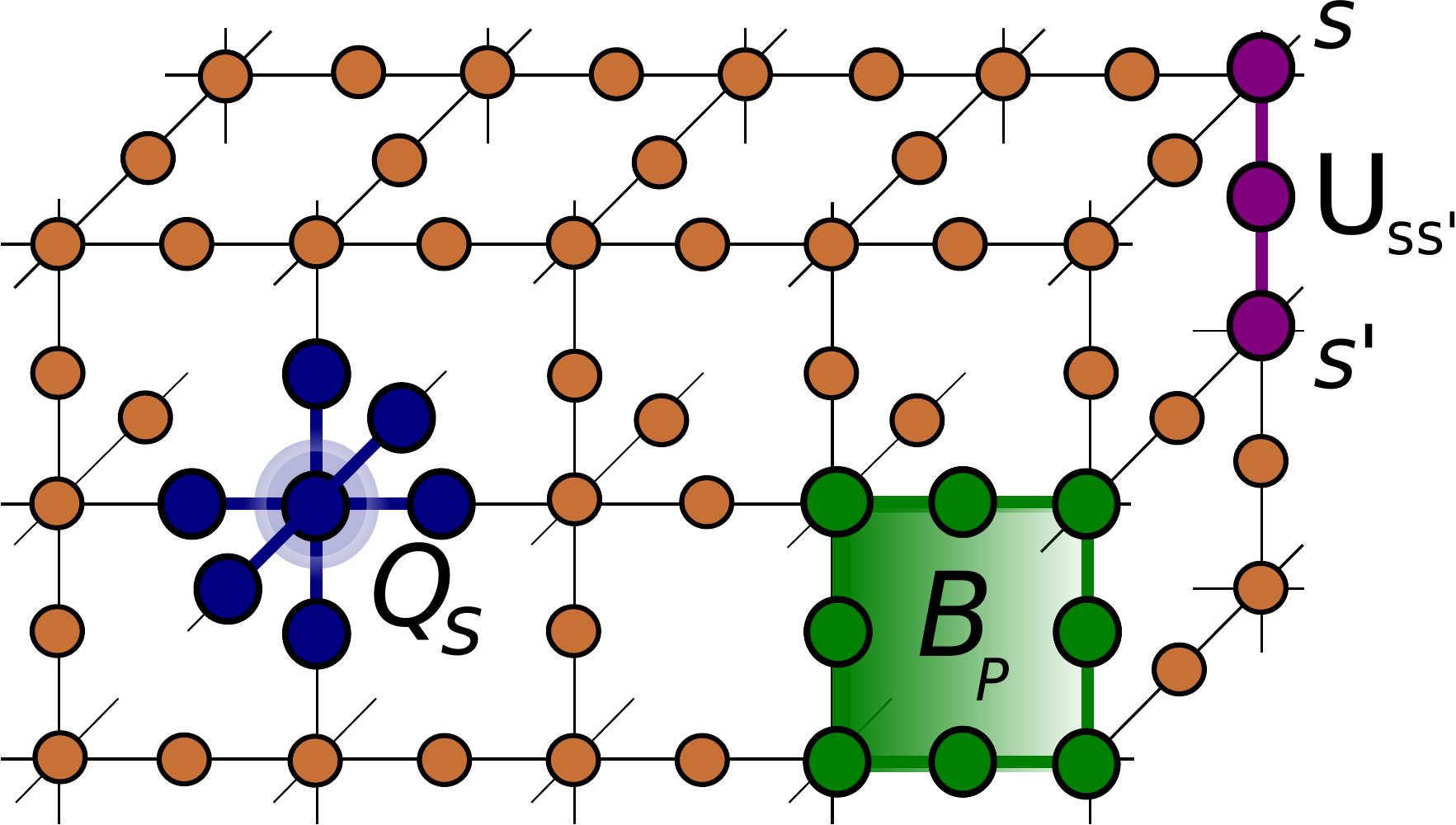}%
	\caption{The bosonic model: the bosons live on a cubic lattice, both on the sites $s$ and on the links $ \braket{ss'}$. The cluster charge $\hat{Q}_{s}$ (\ref{eq:bs1}) measures the charge on the site $s$ and the adjacent link $\braket{ss'}$; the ring exchange $\hat{B}_{p}$ (\ref{eq:bs3}) composed of four hopping operators $\hat{U}_{ss'}$ (\ref{eq:bs4}) acts on all sites and links of a plaquette $p$.
	}
	\label{fig:bosonmodel}
\end{figure}

The bosonic degrees of freedom live on the sites $s$ and links $\braket{ss'}$ of a bipartite cubic lattice $\Lambda$; they carry a $U(1)$ charge of $2e$. Creation and number operators on sites and links  are denoted by $\hat{b}_{s}^{\dagger}$, $
\hat{n}_{s}$  and $\hat{b}_{ss'}^{\dagger}$, $\hat{n}_{ss'}$, respectively. The bosonic Hamiltonian comprises of two terms, which we refer to as the charging and the hopping Hamiltonian:
\begin{align} \label{Bosonic_model}
\hat{H}_{\mathrm{B}} = V \sum_{s} \hat{Q}_{s}^{2} - \frac{u}{2} \sum_{p} \big( \hat{B}_{p} + \hat{B}_{p}^{\dagger} \big),
\end{align}
where $V$ and $u > 0$ are parameters of the model corresponding to the cluster charging and flux energy scales respectively.  
The cluster charge operator $\hat{Q}_{s}$ measures the total charge on site $s$ and the corresponding six links $\braket{ss'}$ of the three-dimensional lattice $\Lambda$ (see Fig.~\ref{fig:bosonmodel}):
\begin{align} 
\hat{Q}_{s} = \alpha_{s} \sum_{s'} \hat{n}_{ss'} + m\hat{n}_{s},
\label{eq:bs1}
\end{align}
with the constant $\alpha_{s} = 1\cdot\mathbbm{1}_A(s)  + (m-1)\cdot\mathbbm{1}_B(s)$, where $\mathbbm{1}_{A/B}(s)$ are the indicator functions for $A/B$ sublattice. The integer $m \geq 2$ is a parameter determining the discrete gauge group $\mathbb{Z}_m$ and, consequently, values of all fractionalized quantities. Since $V > 0$, the charging term is a short-ranged repulsive interaction;  $\hat{B}_{p}$, on the other hand, is a ring exchange term consisting of simultaneous hopping between adjacent sites and links of a plaquette:
\begin{align}\label{eq:bs3}
\hat{B}_{p} = \hat{U}_{s_{1}s_{2}} \hat{U}_{s_{2}s_{3}} \hat{U}_{s_{3}s_{4}} \hat{U}_{s_{4}s_{1}},
\end{align}
where the hopping term on a link is defined as:
\begin{align}\label{eq:bs4}
\hat{U}_{ss'} = \left(\hat{b}_{s}^{\dagger}\right)^{\alpha_{s} - 1} \hat{b}_{s'}^{\dagger} \hat{b}_{ss'}^{\alpha_{s}} + \hat{b}_{s'}^{\alpha_{s'} - 1} \hat{b}_{s} \left( \hat{b}_{ss'}^{\dagger} \right)^{\alpha_{s'}}.
\end{align}
A crucial part of the model are the definitions of the link and site local Hilbert spaces.
The $\hat{b}_{ss'}^{\dagger}$ degrees of freedom on the links are generalized hard-core bosons, whose occupation number $\hat{n}_{ss'}$ is restriced to lie in $\{0,\ldots,m-1\}$. The site bosons, in contrast, are in the rotor representation: $\hat{b}_{s}^{\dagger}\sim e^{i \hat{\theta}_{s}}$, with $[\hat{n}_{s}, \hat{\theta}_{s}] = i$, and consequently their occupation is \emph{integer}-valued: $-\infty \leq n_s \leq \infty$. 

With those definitions in place, the intuitive action of $\hat{B}_{p}$ is to change site occupations by $\pm1 \mod m$ on links around the plaquette, while preserving the total boson number by compensating with hoppings between sites and links.
The operator $\hat{U}_{ss'}$ hops a unit of cluster charge from one end of the link to the other:
\begin{align} \label{eq:bs5}
\big[ \hat{Q}_{r}, \hat{U}_{ss'} \big] = \big( \delta_{rs'} - \delta_{rs}\big) \hat{U}_{ss'}.
\end{align}
Consequently, $\hat{Q}_{s}$ commutes with the product of $\hat{U}_{ss'}$ around any closed loop, which, in particular, implies:
\begin{align} \label{eq:bs6}
\big[ \hat{Q}_{s}, \hat{B}_{p} \big] = 0 \quad 
\forall \ s,p.
\end{align}
 Furthermore, since $\hat{U}_{ss'}^{\dagger} = \hat{U}_{s's} = \hat{U}_{ss'}^{-1}$ we have that $\big[\hat{B}_{p}, \hat{B}_{p'}\big] = \big[\hat{B}_{p}, \hat{B}_{p'}^{\dagger}\big] = 0$. Since also $\big[\hat{Q}_{s}, \hat{Q}_{s'} \big] = 0$, all terms in the Hamiltonian Eq. (\ref{eq:bs1}) commute and thus can be simultaneously diagonalized. 
  The spectra of both operators are discrete: $\hat{Q}_{s}$ trivially, since it counts the number of bosons, and the hopping term $\hat{B}_{p}$ due to the relation $\hat{B}_{p}^m =1$. The eigenvalues of $\hat{B}_{p}$ are consequently given by $b_{p} = e^{\frac{2 \pi i}{m}}$.
 The quantum numbers $q_{s}\in \mathbb Z_{\geq 0}$ and $b_{p}\in \mathbb Z_{m}$ satisfy the constraints: a global $\sum_{s} q_{s} \equiv 0 \left( \text{mod} \ m \right)$ and a set of local constraints $\prod_{p \in \mathcal C} b_{p} = 1$ for every cube $\mathcal{C}$ of the lattice.
Labelling the eigenstates by $\ket{\{ q_{s}, b_{p}\}}$ we obtain the energy spectrum:
\begin{equation} \label{eq:bs10}
E_{\{ q_{s}, b_{p} \}} = V \sum_{s} q_{s}^{2} - \frac{u}{2} \sum_{p} \big( b_{p} + b_{p}^{*} \big).
\end{equation}
This bosonic spectrum is \emph{discrete} even in the thermodynamic limit and hence gapped, with the groundstate $\ket{\mathrm{GS}} = \ket{\{ q_{s} = 0, b_{p} = 1 \}}$. 

There are two types of excitations: charge excitations, created by string operators defined as $\hat{W}(\mathcal{L}) := \prod_{ss' \in \mathcal{L}} \hat{U}_{ss'}$, where $\mathcal{L}$ is any open path/string on the lattice $\Lambda$. Acting on the ground state gives rise to a pair of excitations localized on the two ends of the string [as seen from Eq.~\eqref{eq:bs5}], each costing an energy $V$. 
The second kind are $\mathbb Z_{m}$-flux loops on the dual lattice $\Lambda^{*}$. 
 The topological content of the model \eqref{Bosonic_model} is that of the $\mathbb Z_{m}$ gauge theory \cite{kitaev2003fault}, albeit in an unconventional charge-conserving form which will prove suitable for coupling to a fermionic sector. Another key property is the fact that the bosonic charge excitation is also \emph{fractionally} charged under the $U(1)$ with a charge of $2e/m$ (see SM and \cite{PhysRevB.84.235145}). 
 
To study the interplay of a band-theoretic fermion system with a topological order we couple the bosons in \eqref{Bosonic_model} to additional fermionic degrees of freedom on the sites of $\Lambda$. The total Hilbert space is spanned by the states in the tensor product of the electronic occupation and the bosonic Hilbert space, i.e.,
 $\ket{\left\{n\right\}}_{\mathrm{f}} = \ket{ \{ n_{s \sigma} \}}_{\mathrm{e}} \otimes \ket{ \{ n_{s}, n_{ss'} \}}$. We denote by $\hat{c}_{s \sigma}^{\dagger}$ and $\hat{n}_{s \sigma}$ the creation and number operators for the electrons of spin $\sigma$, and by $\hat{n}_{s,e}$ the total number of electrons on site $s$. The extended Hamiltonian takes the form:
 \begin{align} 
 \hat{H}_{\mathrm{f}} = V \sum_{s}\hat{ \tilde{Q}}_{s}^{2} - \frac{u}{2} \sum_{p} \left(\hat{B}_{p} + \hat{B}_{p}^{\dagger} \right) - \mu \sum_{s \sigma} \hat{n}_{s\sigma},
 \label{Fermionic_model}
 \end{align}
 with $\mu$ the chemical potential and $\hat{ \tilde{Q}}_{s} = \hat{Q}_{s} -  \hat{n}_{s,e} $.
Since all the commutation relations hold with $\hat{Q}_{s}$ replaced by $\hat{\tilde{Q}}_{s}$, the model is diagonalized like before in terms of the eigenstates $\ket{\{ \tilde{q}_{s}, b_{p}, n_{s \sigma}\}}$ with $n_{s \sigma} \in \{0,1\}$. The ground state of the fermionic model is given by: $\ket{\mathrm{GS}}_{\mathrm{f}} = \ket{\{ \tilde{q}_{s} = 0, b_{p} = 1, n_{s \sigma} = 0 \}}$
 when $\mu < 0$. In addition to the charge and flux excitations we also have the spin-$\frac{1}{2}$ fermionic excitation, where
 $n_{s \sigma}$ = 1 for some site $s$ and spin $\{ \sigma = \uparrow , \downarrow \}$. As previously, the electrically charged particles satisfy the global constraint 
 $\sum_{s} \tilde{q}_{s}+ \sum_{s} n_{s,e}  = 0 \; (\text{mod} \; m)$. It is easy to see that in the ground state each electron is accompanied by a bosonic charge to yield together $\tilde{q}_{s} = 0$. Furthermore, it is possible to write a nearest-neighbor hopping Hamiltonian which commutes with $\hat{H}_{\mathrm{f}}$ and moves them together without dissociating:
 \begin{align}  \label{cur2}
 \hat{H}_{\mathrm{hop}} =& - \sum_{\braket{ss'}} \sum_{\sigma \sigma'}\left[ t_{ss', \sigma \sigma'} \hat{c}_{s' \sigma'}^{\dagger} \hat{c}_{s \sigma} \hat{U}_{ss'}  + h.c. \right],
 \end{align}
 where the hopping amplitudes $t_{ss', \sigma \sigma'}$ define the band structure. The boson-electron composites are thus stable and the low-energy excitations are fermions of fractional charge $q_{\mathrm{f}} = e (1 + 2/m)$. In what follows, we choose 
 \begin{equation}
 \begin{split}
 t_{s,s+\hat{\boldsymbol{e}}_\delta}&=i\sigma_\delta-\sigma_z,\qquad \delta=x,y,
 \\
 t_{s,s+\hat{\boldsymbol{e}}_z}&=-\sigma_z,
 \qquad
 t_{s,s}=M\sigma_z,
 \end{split}
 \label{eq: Model Hamiltonian}
 \end{equation}
 with all other $ t_{s,s}=0$ if they are not related by Hermiticity to the ones given. In momentum space, this model has a Weyl semimetal band structure if the real parameter $M\in(-3,3)$. The Bloch Hamiltonian reads
$ \mathcal{H}(\boldsymbol{k}) =\text{sin}(k_{x})\sigma_x+ \text{sin}(k_{y})\sigma_y+[ M - \text{cos}(k_{x})  - \text{cos}(k_{y})  - \text{cos}(k_{z})]\sigma_z$, where $\sigma_{x,y,z}$ are the Pauli matrices acting on spin degrees of freedom. 
 
 \paragraph{Fractional currents --} 
 Charge fractionalization is manifest in the currents. 
 The charge operator on site $s$ (whose sum over all sites, via Eq.~\eqref{eq:bs1}, is the total electric charge of elementary bosons and electrons in the system) is defined as $\hat{\rho}_{s} = \frac{2}{m} \hat{Q}_{s} + \sum_{\sigma} \hat{c}_{s \sigma}^{\dagger} \hat{c}_{s \sigma}$, 
 \begin{align}  
 \hat{\rho}_{s} & = \frac{2}{m} \hat{Q}_{s} + \sum_{\sigma} \hat{c}_{s \sigma}^{\dagger} \hat{c}_{s \sigma},
 \label{eq:charge fermion model}
 \end{align}
 and from the continuity equation for this local quantity:
 \begin{align}  
 \frac{d}{dt} \hat{\rho}_{s} = \frac{i}{\hbar} \left[ \hat{H}, \hat{\rho}_{s} \right] = - \sum_{\delta=x,y,z} \left( \hat{J}_{s + \boldsymbol{\hat{e}}_\delta}^{\delta} - \hat{J}_{s}^{\delta}\right),
 \label{eq:cont eqn}
 \end{align}
 we obtain that the charge current is given by:
 \begin{align} \nonumber
 \hat{J}_{s}^{\delta}=&\frac{i}{\hbar} \frac{m + 2}{m}\sum_{\sigma\sigma'}     \left(t_{(s-\boldsymbol{\hat{e}}_\delta)s, \sigma \sigma'}  \hat{c}_{s \sigma}^{\dagger}  \hat{c}_{(s - \boldsymbol{\hat{e}}_\delta) \sigma'}  \hat{U}_{(s-\boldsymbol{\hat{e}}_\delta)s} - h.c. \right).
 \end{align}
  The total currents $\hat{J}^{\delta}$ are obtained by summing over all sites $s$. Since  $\hat{J}^{\delta}$ commutes with $\small{ \hat{\tilde{Q}}_{s}}$ and $\hat{B}_{p}$, its matrix elements
 in the basis $\ket{\{ \tilde{q}_{s}, b_{p}, n_{s \sigma}\}}$ are diagonal in the $\tilde{q}_{s}$ and $b_{p}$ sectors. We therefore restrict to the low-energy sector $\tilde{q}_{s} = 0$ and $b_{p} = 1$, where the only excitations are fermionic composites of bosonic quasiparticles and electrons. A state in this sector, given by an arbitrary electronic configuration, can be written as $ \ket{ \{ n_{s \sigma} \} }_{\mathrm{f}} = \ket{ \{ \tilde{q}_{s} = 0, b_{p} = 1, n_{s \sigma}\}}$. 
It is easily shown (see SM) that: $\hat{c}_{s \sigma}^{\dagger}  \hat{c}_{(s -\boldsymbol{\hat{e}}_\delta) \sigma'}  \hat{U}_{(s-\boldsymbol{\hat{e}}_\delta)s} \ket{ \{ n_{r \xi} \}}_{\mathrm{f}} = \pm \ket{ \{ n'_{r \xi} \} }_{\mathrm{f}}$, where $n'_{r \xi}  = n_{r \xi} + \delta_{rs} \delta_{\sigma \xi} - \delta_{r(s-\boldsymbol{\hat{e}}_\delta)} \delta_{\sigma' \xi}$, and the sign depends on the ordering of creation operators in the defintion of the electronic basis state. Thus in the low-energy subspace the matrix elements of the hopping/current operator are those of free fermions and we can \emph{define}:
 \begin{align} \label{eq:cur71}
 \hat{d}_{s \sigma}^{\dagger}  \hat{d}_{s' \sigma'} \equiv \hat{c}_{s \sigma}^{\dagger}  \hat{c}_{s'\sigma'}  \hat{U}_{s's},
 \end{align}
 where $\hat{d}_{s \sigma}^{\dagger}$ creates a \emph{fractionally} charged fermionic excitation (that braids non-trivially with the loop-like $\mathbb Z_{m}$ flux excitations when not restricted to low-energy subspace). With this definition the effective quasiparticle current in the low-energy subspace is given by:
 \begin{align}
 \label{eq:cur8}
 \hat{J}^{\delta} = &   \frac{(m + 2)}{m} \frac{i}{\hbar}\sum_{s\sigma \sigma'}\left( t_{(s -\boldsymbol{\hat{e}}_\delta)s, \sigma \sigma'}  \hat{d}_{s \sigma}^{\dagger}  \hat{d}_{(s - \boldsymbol{\hat{e}}_\delta) \sigma'}  - h.c. \right)
 \end{align}
 Having derived the low-energy effective theory, we can consider physical signatures. A standard calculation using the Kubo formula and the current Eq.~(\ref{eq:cur8}) in momentum space yields the conductivity:
\begin{align} \label{eq:apxend20.1}
\sigma_{xy}  = \left(\frac{m+2}{m}\right)^2\frac{e^{2}}{ \pi h } \Delta k  \equiv  \frac{q_{\mathrm{f}}^{2}}{ \pi h } \Delta k.
\end{align}
While having a fractional prefactor, $\sigma_{xy}$ also depends on the distance between the Weyl nodes in the $z$ direction $\Delta k$. It is therefore not universal, and hence cannot serve as an unambiguous signature of the state. Crucially though, there exists a fractional quantized response which \emph{is} universal.
\paragraph{Fractional CPGE --}The part of a photocurrent whose direction switches depending on the circular polarization of the incident light is called circular photogalvanic effect (CPGE) \cite{PhysRevB.61.5337,de2017quantized}. It is given by the second order response:
\begin{align} \label{CPGE}
\frac{d j_{m}}{d t} & =  \beta_{mn}(\omega) \big[ \boldsymbol{E}(\omega) \times \boldsymbol{E}^{*}(\omega)\big]_{n},
\end{align}
with $\boldsymbol{E}(\omega) =\boldsymbol{E}^{*}(- \omega)$ the electrical field. In an unfractionalized Weyl semimetal, the CPGE current is quantized in terms of universal constants $e, h, c, \varepsilon_0$ and the monopole charge of the Weyl nodes, provided they lie at different energies (the assumption holds in enantiomeric crystals, with the inversion and all mirror symmetries broken) \cite{de2017quantized}. We show the CPGE is quantized to a fractional value in our model, and hence provides a sharp diagnostic of the state.

We note that within the solvable model \cite{PhysRevB.84.235145} coupling to the electromagnetic (EM) field is incorporated by introducing vector potentials $A^1_{ss'}$ and $A^2_{ss'}$ on the two halves of the links of the lattice and modifying the hopping operators, in particular $U_{ss'}$, via the Peierls substitution (with phases dependent on charges of elementary bosons/electrons being hopped). For weak and possibly time-dependent fields the resulting low-energy effective theory is shown to be:
\begin{align}  \label{em6}
\hat{H}_{\mathrm{eff}} = & -\sum_{\braket{ss'}} \left( t_{ss', \sigma \sigma'} \hat{d}_{s' \sigma'}^{\dagger} \hat{d}_{s \sigma} e^{i e\left(1 +  \frac{2}{m} \right)A_{ss'}} + h.c. \right) \; \nonumber \\
& - \mu \sum_{s \sigma} \hat{d}_{s\sigma} ^{\dagger} \hat{d}_{s\sigma},
\end{align}
i.e., the fermions $\hat{d}_{s\sigma} ^{\dagger}$ minimally couple to the EM field, with coupling strength given by their fractional charge. Consequently, the tensor $\beta_{mn}(\omega)$ is derived as in Ref.~\cite{PhysRevB.61.5337}, with the elementary electric charge replaced by the fractional $q_\mathrm{f}$. More concretely, in a two-band model \cite{de2017quantized}:
\begin{align} \label{eq:CPGE4}
\beta_{mn}(\omega) = \frac{\pi q_{\mathrm{f}}^{3}}{\hbar^{2} V} \sum_{\textbf{k}}  \left(\partial_{k_{m}}\, \Delta E_{\textbf{k}}\right) \,\Omega_{\textbf{k},n}\, \delta\left( \hbar \omega + \Delta E_{\boldsymbol{k}} \right),
\end{align}
where $V$ is the sample volume, $\Delta E_{\boldsymbol{k}}  $ is the energy difference between valence and conduction band, and $\Omega_{\boldsymbol{k}}$ is the Berry curvature of the valence band.

The delta function in Eq.~(\ref{eq:CPGE4}) selects a k-space surface $\mathcal{S}$, and the trace of the CPGE tensor measures the Berry flux through that surface \cite{de2017quantized}. If, therefore, $\mathcal{S}$ encloses a Weyl node, Eq.~(\ref{eq:CPGE5}) yields the monopole charge of that node, and hence CPGE is quantized \emph{and} fractionalized:
\begin{align} \label{eq:CPGE5}
\text{Tr}[\beta(\omega)] = i \frac{q_{\mathrm{f}}^{3}}{2h^{2}} \oint_{\mathcal{S}} d\boldsymbol{S}  \cdot \boldsymbol{\Omega} = i \pi \frac{q_{\mathrm{f}}^{3}}{h^{2} }C,
\end{align}
where $C$ denotes the monopole charge of the Weyl node. 
\paragraph{Wiedemann-Franz law violation --}
Another signature of fractionalization of the low-energy degrees of freedom is the violation of the Wiedemann-Franz law, which may be anticipated from Eq.~(\ref{eq:cur8}). Indeed, a short computation of thermal and electrical conductivities based on semiclassical Boltzmann approach, in the spirit of Ref.~\cite{PhysRevB.90.121108}, yields:
\begin{align} \label{eq:WFratio}
\frac{\sigma_{xx}}{\kappa_{xx}} = T\frac{\pi^2}{3}\left( \frac{k_{\mathrm{B}}}{q_\mathrm{f}} \right)^2,
\end{align}
where the fractional $q_{\mathrm{f}}$ takes the place of the electron charge. This is akin to what happens to the Hall conductivity in a FQH state. We stress though that our result is valid at zero magnetic field.
\begin{figure}[t]
 	\centering
 	\includegraphics[scale=0.23]{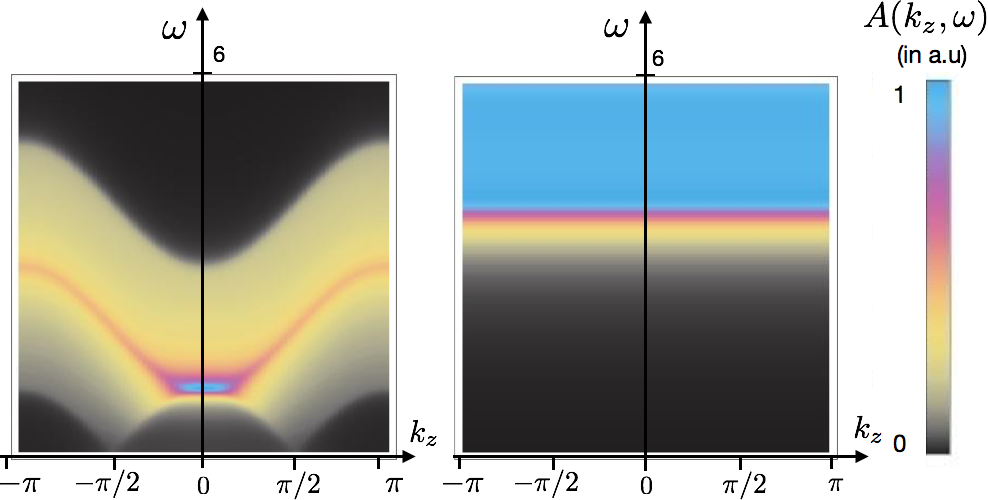}%
 	\caption{The above shows the electron spectral functions numerically summed over $k_{x}$ and $k_{y}$ for a non-interacting Weyl semimetal (left panel) described by \eqref{eq: Model Hamiltonian} and a fractional Weyl-semimetal (right panel), plotted as a function of $\omega$ and $k_{z}$. In the numerical sum, we use the parameters $M=2$, $\eta=.2$, $\mu=0$ and $V=3$. 
 	}
 	\label{fig:spectral}
\end{figure}
\paragraph{Electronic Green's function --}
The low-energy physics is that of Weyl \emph{composites}. What becomes of the electrons? Intuitively, it is clear that they can be created by dissociating the composite, at a finite energy cost. This is indeed reflected in a gap in the electron spectral function $A(r,r',\omega) = 1/\pi \  \mathrm{Im}\ G^{\mathrm{R}}(r,r',\omega)$, where the retarded electronic Green's function: 
\begin{align}
G^R(r,t;r',t') = -i\theta(t-t')\left\langle \left\lbrace  \hat{c}_{r\sigma}(t),\hat{c}^\dagger_{r'\sigma'}(t') \right\rbrace \right\rangle
\end{align}
is evaluated in the ground state of $\hat{H}_\mathrm{f}+\hat{H}_\mathrm{hop}$, and  $\hat{c}_{r\sigma}(t) = e^{i(\hat{H}_\mathrm{f}+\hat{H}_\mathrm{hop})t} \hat{c}_{r\sigma}e^{-i(\hat{H}_\mathrm{f}+\hat{H}_\mathrm{hop})t}$. Evaluation of $\langle \mathrm{GS} | \hat{c}_{r\sigma}^\dagger e^{-i(\hat{H}_\mathrm{f}+\hat{H}_\mathrm{hop})(t-t')}\hat{c}_{r'\sigma'} | \mathrm{GS} \rangle$ is a nontrivial step. Nevertheless, it can be computed explicitly: the key insight is that $|\mathrm{GS}\rangle$ is a superposition, generated by $\hat{H}_\mathrm{hop}$, of electron-boson composites $\hat{d}$, and that $\hat{c}_{r\sigma} | \mathrm{GS} \rangle$ is likewise a superposition of composites with an unpaired boson $q_r$ at site $r$. Furthermore, $e^{-i\hat{H}_\mathrm{hop}t}$ can only move a boson if accompanied by an electron, hence the matrix element vanishes unless $r=r'$, and then it reduces to the free fermion calculation, with electrons replaced by composites $\hat{d}$. Proceeding in the standard fashion we obtain:
\begin{align}\nonumber
A_{\sigma\sigma'}(r,r',\omega) = -\frac{1}{\pi}\mathrm{Im}\left[\delta_{rr'}\delta_{\sigma\sigma'} \sum_{\boldsymbol{k}} \frac{u_{\sigma}^*(\boldsymbol{k})u_{\sigma}(\boldsymbol{k})}{\omega-\xi_{r,\sigma}(\boldsymbol{k})+i\eta}\right],
\end{align}
where $u_{\sigma}(\boldsymbol{k})$ is the Bloch eigenstate of the valence band of Eq.~(\ref{eq: Model Hamiltonian}). Crucially, $A(r,r',\omega)\propto \delta_{rr'}$, which trivializes its momentum dependence (see Fig.~\ref{fig:spectral}). Moreover the effective dispersion, $\xi_{r,\sigma}(\boldsymbol{k})$ is shifted with respect to the free electron dispersion of the valence band $\varepsilon(\boldsymbol{k})-\mu$:
\begin{align}\label{eq:spectral2}
\xi_{r,\sigma}(\boldsymbol{k}) = V(1 - 2q_r+2n_{r,\sigma}) +\varepsilon(\boldsymbol{k})-\mu.
\end{align}
This results in a spectral gap in the spectral function (see Fig.~\ref{fig:spectral}), which is a strong experimental signature. Interestingly, it also suggests an enhanced stability of CPGE to hopping disorder (compared to the noninteracting case).
\paragraph{Discussion -- }
We introduced a solvable lattice model for a fractional Weyl semimetal, and derived experimental signatures of this new phase. Our construction shows that such a phase combining fractionalization with a gapless Weyl bandstructure can be realized in strongly-interacting three-dimensional systems, and allows us to derive distinctive experimental signatures of this phase: gapped electronic quasiparticles, but a fractional CPGE and a fractionally violated Wiedemann-Franz law heralding the gapless nature of the fractional Weyl semimetal.

\paragraph{Acknowledgements -- }
The authors thank Manfred Sigrist and Adolfo Grushin for stimulating discussions.
OT and TM acknowledge financial support from the DFG via the Emmy Noether Programme ME 4844/1-1, SFB 1143 (project-id 247310070), and the Würzburg-Dresden Cluster of Excellence on Complexity and Topology in Quantum Matter - ct.qmat (EXC 2147, project-id 39085490). AS thanks The European Research Council (Project LEGOTOP), The Israeli Science Foundation and the DFG (Project CRC 183) for support.  
MKJ thanks Sebastian Huber for fruitful exchanges. 
TN acknowledges support from the Swiss National Science Foundation (grant number: 200021\_169061) and from the European UnionÕs Horizon 2020 research and innovation program (ERC-StG-Neupert-757867-PARATOP).

\bibliography{fractionalweyl}{}

\begin{thebibliography}{29}
\bibitem [{aba()}]{abanin-misc}%
  \BibitemOpen
  \href@noop {} {}\bibinfo {note} {Reference \cite{PhysRevLett.113.136402} also
  finds an ``orthogonal" strongly interacting Weyl state within slave-particle
  description, however its thermodynamic and transport properties are that of a
  standard Weyl semimetal. Reference \cite{PhysRevB.98.201111} constructs a
  topologically ordered electric insulator with Majorana-Weyl cones of emergent
  fermions, with a (non-quantized) thermal Hall response while
  Reference~\cite{WANG201580} derives a fractionalized response for a
  semi-metal based on a microscopic parton construction.}\BibitemShut {Stop}%
\end{thebibliography}


\section{Supplementary material}
\subsection{Equivalence between the bosonic model and $\mathbb Z_{m}$-topological gauge theory}
In this section we describe the mapping of the bosonic hopping model $\hat{H}_{\mathrm{B}}$ described in the main text (see eq.~\eqref{Bosonic_model}) to a topological $\mathbb Z_{m}$ gauge theory.  Since all terms in $\hat{H}_{\mathrm{B}}$ commute and thus can be simultaneously diagonalized, its eigenstates can be labelled as $\ket{\{ q_{s}, b_{p}\}}$ where
\begin{align} \label{bs9}
\hat{Q}_{s}  \ket{\{ q_{s}, b_{p}\}} &= q_{s}\ket{\{ q_{s}, b_{p}\}}, \; \nonumber \\ 
\hat{B}_{p} \ket{\{ q_{s}, b_{p}\}} &= b_{p}\ket{\{ q_{s}, b_{p}\}}, \; \nonumber \\
\hat{B}_{p}^{\dagger} \ket{\{ q_{s}, b_{p}\}} &= b_{p}^{*}\ket{\{ q_{s}, b_{p}\}},
\end{align}
leading to the eigenvalues
\begin{align} \label{bs10}
E_{\{ q_{s}, b_{p} \}} = V \sum_{s} q_{s}^{2} - \frac{u}{2} \sum_{p} \big( b_{p} + b_{p}^{*} \big),
\end{align}
where $q_{s}\in \mathbb Z_{\geq 0}$ and $b_p\in \mathbb Z_{m}$.  Furthermore the quantum numbers $q_{s}$ and $b_{p}$ satisfy the constraints mentioned in the main text.  The ground state given by $\ket{\mathrm{GS}} = \ket{\{ q_{s} = 0, b_{p} = 1 \}}$ is gapped, which can be immediately inferred from the fact that the eigenvalues of $\hat{Q}_s$ and $\hat{B}_p$ are discrete. 

\medskip In order to illustrate the connection between the bosonic model and a standard $\mathbb Z_{m}$ topological gauge theory which is a 3D $\mathbb Z_{m}$ analog of the 2D $\mathbb Z_2$ toric code \cite{kitaev2003fault}, next we study the excitations and correlation functions of $H_{\mathrm{B}}$. As in a discrete gauge theory there are two kinds of excitations within this model, i.e charge excitations and flux-excitations. 
\begin{itemize}
\item {\it{Charge excitations:}} These are fractionalized point-like excitations that are non-local in the sense that they have a string attached to them. They are created in pairs via the action of the string operators defined as $\hat{W}(\mathcal{L}) := \prod_{ss' \in \mathcal{L}} \hat{U}_{ss'}$ on the groundstate, where $\mathcal{L}$ is any open path/string on the lattice $\Lambda$. Acting on the ground state gives rise to a pair of excitations localized on the two ends of the string, each costing an energy $V$. More generally the energy of a state $\hat{W}(\mathcal{L}) \ket{\{ q_{s}, b_{p} \} }$ is given by $E_{ \{ q_{s}, b_{p} \}} + 2V \big[ 1 +  q_{s_{f}}-q_{s_i}]$ where $s_{f}$ and $s_{i}$ are the end point and initial point respectively of the oriented line $\mathcal L$.
\item {\it{Flux-loop excitations:}} The second kind of excitations are $\mathbb Z_{m}$-flux loops on the dual lattice $\Lambda^{*}$. Consider a surface $\mathcal S^{*}$ which is a product of `dual' plaquettes $p^{*}\in \mathcal S^{*}$. Since there is a one-to-one correspondence between plaquettes in $\Lambda^{*}$ and links in $\Lambda$, we may replace the product over plaquettes $\prod_{p^{*}\in \mathcal S^{*}}$ by $ \prod_{ ss' \in \mathcal S^{*}}$. Then, a flux-loop excitation can be created by the operator $\hat{O}(\mathcal S^{*}) := \prod_{ ss' \in \mathcal S^{*}}e^{\frac{2 \pi i }{m} \cdot \hat{n}_{ss'}}$, where $\hat{n}_{ss'}$ are the bosonic link operators. The state $\hat{O}_{ss'} \ket{\text{GS} }$ carries a flux-loop on the boundary of $\mathcal S^{*}$. This can be seen by evaluating the eigenvalues of $\hat{B}_{p}$ for this state 
\begin{align}
\hat{B}_p \hat{O}_{ss'} \ket{\mathrm{GS}}= 
e^{\frac{2\pi i\text{Int}(p,\partial S^{*})}{m}}\hat{O}_{ss'} \ket{\mathrm{GS}}, 
\end{align}
where $\mathrm{Int}(p,\partial S^{*})$ is 1 if the boundary of $\mathcal S^{*}$  intersects the plaquette $p$ and 0 otherwise.
\end{itemize}

\medskip The charge excitations acquire an Aharonov-Bohm phase of $\theta = \frac{2 \pi}{m}$ when braided around a flux excitation.  The braiding can be seen by moving a charge excitation along a loop $\gamma$ (by a sequence of hopping operators $\hat{U}_{ss'}$, with $ss'\in \gamma$) such that $\gamma$ links non-trivially with the flux operator which bounds an open surface on the dual lattice $\Lambda^{*}$. 

\medskip So far we have implicitly only considered a spatial lattice with trivial topology such that all closed loops and closed surfaces are contractible. Next, we consider putting the model (\ref{Bosonic_model}) on a lattice with non-trivial topology. A canonical choice is the three-torus $T^{3}$ which is achieved by simply imposing periodic boundary conditions in all three directions on a cubic lattice. The three-torus has three non-contractible 1-cycles and three non-contractible 2-cycles. Let us label these $\Lambda_{i}$ and $\Sigma_{i}^{*}$. These cycles are the generators of the homology groups $H_{1}(T^{3},\mathbb Z)$ and $H_{2}(T^{3},\mathbb Z)$. Then we may define the operators 
\begin{align} \label{bs16}
\hat{W}\left( \Lambda_{i} \right) & := \prod_{ss' \in \Lambda_{i}} \hat{U}_{ss'},  \nonumber \\
\hat{O}(\Sigma^{*}_{i}) & = \prod_{{ss' \in \Sigma^{*}_i }} e^{\frac{2 \pi i }{m} \cdot \hat{n}_{ss'}}.
\end{align}
These operators commute with the Hamiltonian \eqref{Bosonic_model} and satisfy the algebraic relations 
\begin{align} 
\hat{W}\left( \Lambda_{i} \right) \hat{O}(\Sigma^{*}_{i'}) = e^{ \frac{2 \pi i}{m} \delta_{ii'}} \hat{O}(\Sigma^{*}_i)\hat{W}\left( \Lambda_{i} \right).
\label{GS_algebra}
\end{align}
Therefore the groundstate Hilbert space must furnish a representation of the algebra \eqref{GS_algebra}. A minimal irreducible representation has dimension $m^{3}$ and can be labelled as
\begin{align} \label{bs16.2}
\ket{\{ n^{x}, n^{y}, n^{z} \}} \in \mathcal{H}^{\mathrm{GS}}_{T^{3}}.
\end{align}
where $n_{i} \in [0,1,\cdots,m-1]$, which satisfy
\begin{align}
\hat{W}\left( \Lambda_{x} \right)|n_{x},n_y,n_{z}\rangle =&\; e^{\frac{2\pi in_x}{m}}|n_x,n_y,n_z \rangle, \nonumber \\
\hat{O}_{x}\left( \Sigma_{x}^{*} \right)|n_{x},n_y,n_{z}\rangle =&\; |n_x+1 \ \text{mod} \ m,n_y,n_z \rangle.
\end{align}
and analogously for $i=y,z$. $\Sigma^{*}_{i}$ should be thought of as the 2-cycle orthogonal to the 1-cycle $\Lambda_i$. To summarize, we have shown that (i) point-like charge excitations and loop-like flux excitations have non-trivial Aharonov-Bohm like topological correlations and (ii) the ground state degeneracy is sensitive to the topology of the underlying lattice the model is defined on. In fact these are the defining properties of  the $\mathbb Z_{m}$ topological gauge theory \cite{kitaev2003fault}.

\subsection{Heat current and fractional Wiedemann-Franz law }
Let us now define the thermal current operators associated to a fractional Weyl semimetal. As shown above, the charge current operator of the fractional Weyl semimetal is of the same form as that of a current operator in a ``regular'' (non-fractionalized) Weyl semimetal but with a fractional multiplicative factor fixed by the charge of the fermionic excitations in the low energy subspace of the model. An energy current, however, is naturally insensitive to the charge of the quasiparticles carrying this energy current. We now show that the energy current associated with our fractional Weyl semimetal has the same form as that of a regular Weyl semimetal.\\~\\%
To derive the form of the energy current, we use the fact that energy is conserved and consequently, the energy current satisfies a continuity equation. Following exactly along the lines of Ref.~\onlinecite{vcg2006}, we start out by defining the local energy density on site $s$ as all local terms in the tight-binding Hamiltonian associated with site $s$, plus $1/2$ times all hopping terms connecting this site to other sites. We call the sum of these terms $h_s$, such that $H=\sum_s h_s$. The continuity equation then states for a site $s$ that
\begin{equation}
	\frac{dh_s}{dt}=i[H,h_s]=-\sum_{\boldsymbol{\alpha}}j^{E,\boldsymbol{\alpha}}_{s+\boldsymbol{\alpha}}-j^{E,\boldsymbol{\alpha}}_{s},
\end{equation}
where \begin{equation}
h_{s}=\frac{1}{2}\sum_{\sigma,\sigma',\boldsymbol{\delta}}\left[t_{s,s+\boldsymbol{\delta},\sigma,\sigma'}d^{\dagger}_{s+\boldsymbol{\delta},\sigma'}d^{\phantom{\dagger}}_{s,\sigma}+h.c.+h_{\rm bos}\right],
\end{equation} 
is the local energy density with $\delta=(\hat{x},\hat{y},\hat{z})$ and $h_{\rm bos}$ is the energy density in the bosonic subsystem (which commutes with hopping).. The energy current operator can now be extracted as
\begin{equation}{\label{eq:current}}
	j^{E,\boldsymbol{\alpha}}_{s}=-\frac{i}{2}\sum_{\sigma,\sigma',\boldsymbol{\delta}}\left[t_{s,s+\boldsymbol{\delta},\sigma,\sigma'}\,t_{s-\boldsymbol{\alpha},s,\sigma',\sigma}\,d^{\dagger}_{s+\boldsymbol{\delta},\sigma'}d^{\phantom{\dagger}}_{s-\boldsymbol{\alpha},\sigma'}+h.c.\right],
\end{equation}
where $\alpha=(\hat{x},\hat{y},\hat{z})$. We thus find that the energy current operator is indeed of the exact form as the energy current of a regular Weyl semimetal. Let us now discuss the thermal and electrical conductivity associated with the above-defined fractional Weyl semimetals based on a Boltzmann approach. We stress that we focus on the case of zero magnetic field, and follow Ref.~\cite{k2014}. In the following, we adopt a low-energy picture of a system with two Weyl nodes only, one of positive chirality $\chi=+1$, and one of negative chirality $\chi=-1$. We furthermore assume the chemical potential to be close to the nodes, such that the low-energy excitations live on two distinct Fermi surfaces that correspond to spheres around the individual Weyl nodes. For the sake of the argument, we neglect finer effects such as the orbital magnetic moment or similar: they do not affect our main conclusion below.\\~\\
The semi-classical equations of motion of Weyl fermions subject to only external electric field are well-known 
\begin{align}
	\dot{\boldsymbol{r}}_{\pm}&=\boldsymbol{v_k}+q_{\mathrm{f}}\boldsymbol{E}\times\boldsymbol{\Omega(k)}^{\pm},\\
	\dot{\boldsymbol{k}}_{\pm}&=q_{\mathrm{f}}\boldsymbol{E},
\end{align}  
where $\Omega^\pm(k)$ is Berry curvature on the Fermi surface around the node of chirlaity $\chi$, $v(k)$ is the Fermi velocity, and $\boldsymbol{E}$ is the applied electric field. The Boltzmann equation with the relaxation time approximation is given as 
\begin{equation}
	\frac{\partial f^\pm}{\partial t}+\boldsymbol{v}_\pm\cdot\nabla_{\boldsymbol{r}} f^\pm+\boldsymbol{F}\cdot\nabla_{\boldsymbol{k}} f^\pm=\bigg(\frac{\partial f^\pm}{\partial t}\bigg)_{\text{colission}}m\label{eq:boltzmann}
\end{equation}
where $f^\pm=f^\pm(k,r,t)$ is non-equilibrium distribution function on the Fermi surface of chirality $\chi=\pm$, $\boldsymbol{v}_\pm=\dot{\boldsymbol{r}}_{\pm}$ is the associated velocity, and $\boldsymbol{F}=\dot{\boldsymbol{k}}_{\pm}$ the force. The right-hand side of the equation corresponds to the collision integral that encodes both intra- and inter-Fermi surface relaxation processes. Plugging the semi-classical equations of motions into the Boltzmann equation \eqref{eq:boltzmann}, specializing to the steady state, and keeping only the leading corrections to the equilibrium distribution function $f_{\rm eq}$ yields

\begin{align}
&{\notag}\boldsymbol{v_k}\cdot\bigg(\boldsymbol{\nabla_r}\mu^{\pm}+(\epsilon_k-\mu^\pm)\frac{\nabla_r T}{T}\bigg)\bigg(-\frac{\partial f_{\rm eq}(\epsilon_k)}{\partial \epsilon_k}\bigg)\nonumber\\
&+q_{\mathrm{f}}\boldsymbol{E}\cdot\boldsymbol{\nabla_k}f^{\pm}(k)\nonumber\\
=&\Gamma[f_{\pm}(\boldsymbol{k})-f_{\rm eq}(\boldsymbol{k})]-\Gamma'[f_{\pm}(\boldsymbol{k})-f_{\mp}(\boldsymbol{k})],
\end{align}
where $f^{\pm}$ is distrubtion function for each chiral Fermi surface, $\Gamma$ is scattering rate within the chiral fermi surfaces, $\Gamma'$ denotes the scattering rate between Fermi surfaces, and $\mu^\pm$ is the chemical potential on a chiral Fermi surface. The thermal and electric currents are given as
\begin{align}
	\boldsymbol{j}^\pm_{\mathrm{el}}&=q_{\mathrm{f}}\int \frac{d^3\boldsymbol{k}}{2\pi^3}(\boldsymbol{v_k}+q_{\mathrm{f}}\boldsymbol{E}\times\boldsymbol{\Omega(k)}^{\pm})f^{\pm}(\boldsymbol{k}),\\
	\boldsymbol{j}^\pm_{\mathrm{th}}&=\int \frac{d^3\boldsymbol{k}}{2\pi^3}(\epsilon(\boldsymbol{k})-\mu^\pm)(\boldsymbol{v_k}+q_{\mathrm{f}}\boldsymbol{E}\times\boldsymbol{\Omega(k)}^{\pm})f^{\pm}(\boldsymbol{k}).
\end{align}
Solving the Boltzmann equation as in Ref.~\onlinecite{k2014}, we obtain
\begin{widetext}
 \begin{align}
j^\pm_{x,\mathrm{el}}&=q_{\mathrm{f}}\int \frac{d^3\boldsymbol{k}}{2\pi^3}\bigg(-\frac{\partial f_{\rm eq}(\epsilon_k)}{\partial \epsilon_k}\bigg)v_x^2\bigg(1+\frac{\Gamma'}{\Gamma}\bigg)\frac{1}{\Gamma}\bigg[q_{\mathrm{f}}\bigg(E-\frac{1}{q_{\mathrm{f}}}\partial_x\mu^\pm\bigg)-\big(\epsilon_k-\mu^{\pm}\big)\frac{\partial_xT}{T}\bigg]\\
j^\pm_{x,\mathrm{th}}&=\int \frac{d^3\boldsymbol{k}}{2\pi^3}\bigg(-\frac{\partial f_{\rm eq}(\epsilon_k)}{\partial \epsilon_k}\bigg)(\epsilon_k-\mu^{\pm})v_x^2\bigg(1+\frac{\Gamma'}{\Gamma}\bigg)\frac{1}{\Gamma}\bigg[q_{\mathrm{f}}\bigg(E-\frac{1}{q_{\mathrm{f}}}\partial_x\mu^\pm\bigg)-(\epsilon_k-\mu^{\pm})\frac{\partial_xT}{T}\bigg].
\end{align}
\end{widetext}
Thermal and electric conductivities are then given by
\begin{align}
	\sigma_{xx}&=\sigma_0\,\bigg(1+\frac{\Gamma'}{\Gamma}\bigg),\\
	\kappa_{xx}&=T\,\sigma_0\,\frac{\pi^2}{3}\,\bigg(\frac{k_B}{q_{\mathrm{f}}}\bigg)^2\bigg(1+\frac{\Gamma'}{\Gamma}\bigg),
\end{align}
where $\sigma_0$ is residual electrical conductivity, and $k_B$ is the Boltzmann constant. While each of the two conductivities is non-universal due to their dependence on the intra- and inter-Fermi surface scattering rates $\Gamma$ and $\Gamma'$, their ratio is given by the universal value
\begin{equation}
	\frac{\sigma_{xx}}{\kappa_{xx}}=T\frac{\pi^2}{3}\bigg(\frac{k_B}{q_f}\bigg)^2.
\end{equation}
For a regular Weyl semimetal, the charge $q_{\mathrm{f}}$ would equal the electron charge, $q_{\mathrm{f}}=e$, and the ratio of electrical and thermal conductivities would satisfy the Wiedemann-Franz law. This is clear contrast with the fractional Weyl semimetal, where $q_{\mathrm{f}}=e\,(1+2/m)$. The standard Wiedemann-Franz law is then violated, but a generalized Wiedemann-Franz law involving the fractional charge $q_{\mathrm{f}}$ still holds. This fractional Wiedemann-Franz law at zero magnetic thus provides an experimental fingerprint  of the fractional nature of our system.

\end{document}